\documentclass{elsart4}
\usepackage{graphics}
\usepackage{graphicx}
\usepackage{amssymb}
\usepackage{amsbsy}
\begin{document}

\begin{frontmatter}

\journal{Physica B (proceedings of SCES'05)}
\date{6 July 2005}

\title{Magnetocaloric effect in two-dimensional spin-1/2 antiferromagnets}

\author[BS]{A.~Honecker\corauthref{1}},
\ead{a.honecker@tu-bs.de}
\author[ST]{S.~Wessel}

\address[BS]{Technische Universit\"at Braunschweig,
Institut f\"ur Theoretische Physik,
Mendelssohnstrasse 3, 38106 Braunschweig, Germany}
\address[ST]{
Universit\"at Stuttgart,
Institut f\"ur Theoretische Physik III,
Pfaffenwaldring 57, 70550 Stuttgart, Germany}

\corauth[1]{Corresponding Author.
Tel: +49-531-391 5190,
fax: +49-531-391 5833}

\begin{abstract}

The magnetocaloric effect is studied at
the transition to saturation in the antiferromagnetic spin-1/2 Heisenberg
model on the simplest two-dimensional lattices, namely the square and the
triangular lattice. Numerical results are presented for the entropy which
are consistent with identical universal
properties. However, the absolute values of the entropy
are bigger on the geometrically frustrated triangular lattice than on
the non-frustrated square lattice, indicating that frustration
improves the magnetocaloric properties.

\end{abstract}

\end{frontmatter}

Recent theoretical investigations predict an enhanced magnetocaloric effect
in geometrically frustrated spin systems at low temperatures \cite{Zhito,ZhiHo}.
Indeed, experiments on the pyrochlore magnet Gd$_2$Ti$_2$O$_7$
found substantial drops of temperature around the saturation field during
adiabatic demagnetization \cite{SPSGBPBZ}. These results suggest applications
for efficient low-temperature magnetic refrigeration.

Investigations of quantum spin systems on two-dimensional
lattices have revealed a plethora of semi-classically ordered
and unusual quantum ground states (see \cite{lhuillier03,RSH04}
for recent reviews). Rich behavior is also found in a magnetic field,
including plateaux in the zero-temperature magnetization curve
and field-induced quantum phase transitions (compare \cite{RSH04,comp2D}
and references therein).

However, the magneto-thermodynamics on
such lattices has received suprisingly little attention. As a first
step we study the quantum phase transition at the saturation field
on the square and triangular lattice. The Hamiltonian for a
Heisenberg antiferromagnet in an external magnetic field $h$ reads
\begin{equation}
H = J \sum_{\langle i,j \rangle} \vec{S}_i \cdot \vec{S}_j
 - h \sum_{i} S^z_{i} \, .
\label{Hheis}
\end{equation}
The $\vec{S}_i$ are spin $s=1/2$ operators at site $i$,
$J > 0$ is the exchange constant,
and $\langle i, j \rangle$ are nearest
neighbor pairs on the given lattice.
The magnetocaloric properties are determined by the
entropy $S(h,T)$ at temperature $T$.
One approach is to compute $S(h,T)$ from the
eigenvalues of the Hamiltonian (\ref{Hheis}) \cite{ZhiHo} which are
obtained by exact diagonalization (ED) on a finite lattice with $N$ sites.
Complete spectra are obtained for $N \le 21$ while for bigger
$N$ we use a truncation procedure valid at low energies \cite{HoWe}.
Another approach is to obtain the entropy from a stochastic evaluation of
the partition function using an extended ensemble quantum Monte Carlo
(QMC) method,
based on a flat-histogram formulation~\cite{twa}.
The latter method is restricted to unfrustrated lattices, 
and we employ it for the square lattice.

\begin{figure}[!t]
\begin{center}
\includegraphics[width=\columnwidth]{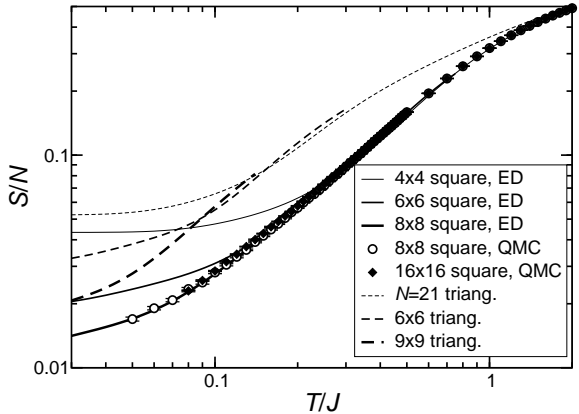}
\end{center}
\caption{Entropy per site of the antiferromagnetic $s=1/2$ Heisenberg
model at the saturation field on the square and the triangular (triang.)
lattice with $N$ sites as a function of temperature $T$. Note the doubly
logarithmic scale.}
\label{SThsat}
\end{figure}

Fig.~\ref{SThsat} shows the entropy per site $S/N$ as a function
of temperature at the
saturation field, given by $h_{\rm sat.} = 4 J$ and $9 J/2$ for
the square and triangular lattice, respectively (see e.g.\
\cite{comp2D}). This corresponds to a quantum critical point
in dimension $d=2$ with a dynamical critical exponent $z=2$.
According to a scaling theory for the low-temperature physics at such
a quantum critical point \cite{ZGRS}, the entropy should follow
a power law
$S(h_{\rm sat.},T)/N \propto T^{d/z} = T$
with possible logarithmic corrections characteristic for $d=2$.
Finite-size effects are evident at low temperatures in
Fig.~\ref{SThsat}, but at intermediate temperatures the curves collapse
onto a line which is consistent with the above univeral power law.
However, the prefactor is clearly different, {\it i.e.}, at low temperatures
the entropy for the frustrated triangular lattice is roughly twice as big
as for the non-frustrated square lattice.

\begin{figure}[!t]
\begin{center}
\includegraphics[width=\columnwidth]{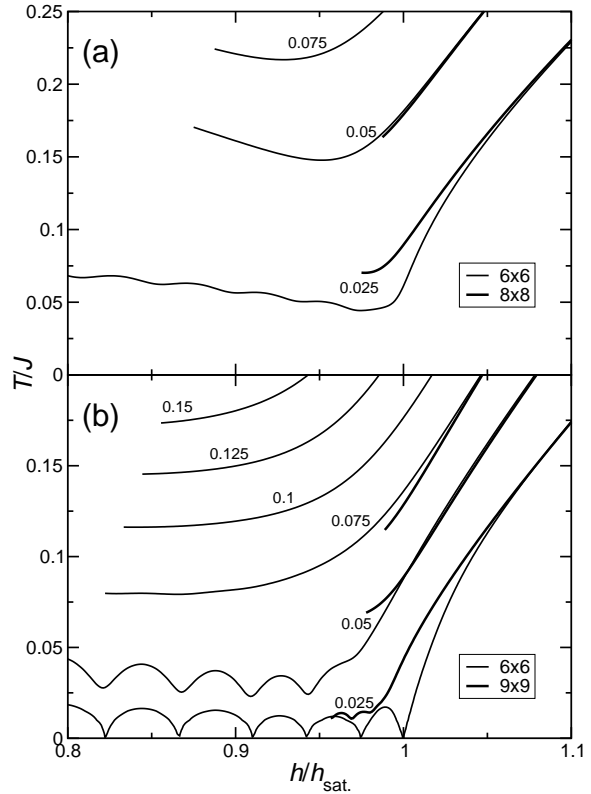}
\end{center}
\caption{Curves of constant entropy on the
(a) square and (b) triangular lattice in the vicinity
of the saturation field $h_{\rm sat.}$. The values of $S/N$
are indicated by numbers next to the lines.}
\label{cEnt}
\end{figure}

Fig.~\ref{cEnt} shows curves of constant entropy in the $h$-$T$
plane close to the saturation field for (a) the square and
(b) the triangular lattice. Since the condition for an adiabatic
process is that entropy remains constant, these curves present
the behavior of the
temperature $T(h)$ during an adiabatic demagnetization process.
Finite-size effects are relevant on the $6 \times 6$ lattices
for $T \lesssim 0.1 \, J$
(the wiggles at very low $T$ correspond to the different
discrete sectors of total $S^z$ at the given system size).
Nevertheless, a drop in temperature is evident when the saturation
field $h_{\rm sat.}$ is approached adiabatically from above in
both cases. There seems to be a slight heating when the field
is further lowered ($h < h_{\rm sat.}$) for the square
lattice (Fig.~\ref{cEnt}(a)) while temperature stays almost
constant on the triangular lattice (Fig.~\ref{cEnt}(b)).
Note furthermore that in the region covered by Fig.~\ref{cEnt},
the values of the entropy $S/N$ are roughly twice as big on the
triangular lattice when compared with the corresponding
region of the square lattice, as we have already observed
for $h = h_{\rm sat.}$.

To summarize, the same universal behavior of $S(T)$ is found on the
square and the triangular lattice at the saturation field.
However, the prefactors are enhanced by
frustration, thus improving the magnetocaloric properties of the
frustrated triangular lattice as compared to the square
lattice, a non-frustrated lattice. Entirely different
behavior is expected for highly frustrated lattices,
where at the classical level cooling rates have been found
to be up to several orders of magnitude bigger \cite{Zhito}.
The magneto-thermodynamics of the corresponding quantum systems
remains to be investigated.

We would like to thank J.\ Richter and M.E.\ Zhitomirsky for useful discussions
and the computing center of the TU Braunschweig for allocation of CPU time on
{\tt cfgauss}.

\end{document}